\documentclass[preprint,times,authoryear]{elsarticle}
\usepackage{amssymb}
\usepackage{lipsum}
\usepackage{amsmath}
\usepackage[colorlinks=true,citecolor=blue,pdfauthor=author]{hyperref}
\usepackage{graphicx,graphics,color}
\usepackage{ulem}
\usepackage{color}
\usepackage{lineno}
\usepackage[T1]{fontenc}
\usepackage{orcidlink}

\newcommand{\arcsec}{\ensuremath{^{\prime\prime}}}

\def\HI{{\rm H\,{\textsc{\romannumeral 1}}}}

\journal{New Astronomy}

\begin{document}

\begin{frontmatter}

\title{Properties of cluster red-sequence spiral galaxies}

\author[a]{Wayne A. Barkhouse\corref{cor1}\orcidlink{0000-0001-5547-3938}}
\ead{wayne.barkhouse@und.edu}
\affiliation[a]{organization={Department of Physics and Astrophysics, University of North Dakota},
            addressline={101 Cornell Street}, 
            city={Grand Forks},
            postcode={58202}, 
            state={North Dakota},
            country={USA}}

\author[b]{Lane M. Kashur\orcidlink{0000-0001-7811-1286}}
\ead{Lane.Kashur@colostate.edu}
\affiliation[b]{organization={Department of Physics, Colorado State University},
            addressline={200 West Lake Street}, 
            city={Fort Collins},
            postcode={80523-1875}, 
            state={Colorado},
            country={USA}}

\author[a]{Moreom Akter\orcidlink{0000-0003-1825-1214}}
\ead{moreomakter94@gmail.com}
\author[a]{Sandanuwan P. Kalawila\orcidlink{0000-0003-1981-8569}}
\ead{kvsprasadh@gmail.com}

\author[c]{Gihan L. Gamage\orcidlink{0009-0009-3329-7596}}
\ead{glgamage@nmsu.edu}
\affiliation[c]{organization={Arts and Science Division, New Mexico State University - Alamogordo},
            city={Alamogordo},
            postcode={88310},
            state={New Mexico},
            country={USA}}

\author[d]{Omar L\'opez-Cruz\orcidlink{0000-0002-1381-7437}}
\ead{omarlx@inaoep.mx}
\affiliation[d]{organization={INAOE},
            city={Tonantzintla},
            postcode={72840},
            state={Puebla},
            country={Mexico}}

\cortext[cor1]{Corresponding author}

\begin{abstract}

We identify a sample of 324 red and 273 blue face-on spiral galaxies selected from 115 low-redshift ($0.014 < z < 0.18$) galaxy clusters imaged with CFHT+MegaCam in $u$- and $r$-band, KPNO 0.9-meter 2TKA and MOSAIC 8K camera in $B$ and $R_{c}$, and images and catalogs extracted from the WINGS survey. Multi-wavelength photometry and spectroscopy were obtained by cross-matching sources with SDSS, {\it GALEX}, and {\it WISE} datasets. Primary findings indicate that up to 45\% of optically red spiral galaxies exhibit significant dust content compared to blue spiral galaxies, as determined by infrared observations. This dust obscuration can conceal ongoing star formation, which may result in the missclassification of red spirals as passively evolving systems. Conversely, approximately half of the red spirals lack substantial dust abundance and appear optically red due to passive evolutionary processes. Support for the passive nature of these red spirals is provided by SDSS emission line data based on the D\textsubscript{n}(4000) spectral index, EW(H$\alpha$), EW(H$\delta$), and [O III] $5007\,\text{\AA}$ luminosity, and on a comparison of the star formation rate and the specific star formation rate with cluster blue spirals. Red spirals are an important link in the evolution of galaxies in the high-density cluster environment and play a key role in determining the physical mechanisms that are responsible for transforming blue star-forming galaxies into red spiral systems.
\end{abstract}

\begin{keyword}
\texttt{}galaxies: clusters: general -- galaxies: spiral -- galaxies: star formation
\end{keyword}
\end{frontmatter}

\section{Introduction}
\label{sect:intro}

A fundamental goal in the study of the galaxy population in clusters is to understand how individual galaxies evolve with time, and to explore what impact the high-density cluster environment has on galaxy evolution \citep[cf.][]{Barkhouse2009,Rude2020,Boselli2022}. It is well established that galaxy type is correlated with local density, as evident by the morphology-density relation \citep{Dressler1980,Dressler1997,Goto2003b}. Elliptical and S0 galaxies form the largest fraction of massive galaxies in the central regions of clusters, while blue spirals dominate in the outskirts. It is important to understand the color and morphology changes that galaxies experience as they enter the cluster environment. For example, the morphology-density relation implies that blue spiral galaxies are transformed into red early-type systems, but where does this transition occur, and what mechanisms are responsible \citep[e.g.,][]{vanderWel2010,Rizzo2018,Pfeffer2023,Vulcani2023}? 

A major clue to the transition of blue spirals into red early-type galaxies was the recognition by \citet{vandenbergh1976} of anemic spirals in galaxy clusters. Since these spirals are characterized by redder than average optical color compared to field spirals, it was assumed that red anemic spirals represent a phase in the evolution of blue spirals into S0s, consistent with the morphology-density relationship \citep{Dressler1984,Bekki2002,Rizzo2018}. Cluster spirals, on average, have a lower specific star formation rate and are depleted in \HI\ compared to spirals located in low-density regions \citep[e.g.,][]{Giovanelli1981,Elmegreen2002,Boselli2006,Denes2016}.

Numerous studies show that star formation in galaxies quenches as they enter high-density cluster environments \citep[see, for example,][]{Wolf2009,Mahajan2012,Bekki2014,Taranu2014,Hamabata2019}. If quenching occurs over a long enough period, galaxies should be observed in a transition state between “blue” and “red” systems, as delineated by their position in the cluster color-magnitude diagram \citep[CMD; ][]{Martin2007,Salim2012}. These “green valley” galaxies (those on the CMD that are located between red and blue galaxy populations) may be a population of transitioning galaxies \citep[e.g.,][]{Wyder2007,Schawinski2014,Smethurst2015,Smethurst2017,Belfiore2018,Blank2022,Vicente2024}.

An important constraint on any process that affects the evolution of spiral galaxies in high density regions is the observed presence of red spiral galaxies in clusters \citep[e.g., ][]{Bekki2002,Goto2003a,Moran2006,Wolf2009,Masters2010,Fraser-McKelvie2018,Cui2024}. Simulations by \citet{Bekki2002} indicate that once star formation ends, spiral arms can remain for a few Gyrs before fading away. This result supports the view that if a dynamical process, such as ram-pressure stripping or galaxy-galaxy interaction, acts to suppress star formation, spiral arms may remain for a few Gyrs \citep{Goto2003a}. After quenching, blue spiral galaxies would then passively evolve and turn red in color, eventually matching the properties of red spirals \citep[e.g.,][]{Bekki2002,Boselli2006,Masters2010,Cerulo2017,Pak2019}.

Some have suggested that red spiral galaxies have a higher dust content than average and appear red in optical colors due mainly to extinction and not necessarily due to the cessation of star formation \citep{Wolf2009,Bosch2013}, although others oppose this suggestion \citep{Masters2010,Rowlands2012}. It has also been stated that red spirals are not passive systems, and that star formation is ongoing but not well-measured by optical photometry \citep{Cortese2012,Cortese2020}.

For this study, we are interested in understanding the transition of blue spirals into red spirals in high-density galaxy cluster environments. We have compiled a sample of red and blue face-on spiral galaxies selected in a well-defined manner from 115 low-redshift ($0.014<z<0.18$) galaxy clusters. Our goal is to identify and quantify the dominant physical mechanism that is responsible for transforming blue cluster spirals into red systems, while preserving spiral arm structure.

The concordance cosmological parameters of $\mathrm{H_0}=70~\mbox{km}~\mbox{s}^{-1}~\mbox{Mpc}^{-1}$, $\Omega_{\Lambda}=0.7$, and $\Omega_{\text{M}}=0.3$ are used for this study \citep{Aghanim2020}. Given the low redshift range of our galaxy sample ($0.014<z<0.18$), our results are not overly sensitive to our adopted cosmology. 

\section{Data and sample selection}\label{sec:data}

Our data consists of $u$- and $r$-band observations of 14 Abell galaxy clusters obtained with the 3.6-meter Canada-France-Hawaii Telescope (CFHT) using MegaCam. Details regarding image processing and photometry can be found in \citet{Rude2020}. Supplementing the CFHT sample are $B$- and $R_{c}$-band images of 31 Abell clusters observed with the KPNO 0.9-meter telescope. These observations were acquired using the T2KA detector and the MOSAIC 8k camera, and are fully described in \citet{LopezCruz2004} and \citet{Barkhouse2007}. Finally, we include 70 galaxy clusters from the WINGS survey \citep{Fasano2006,Varela2009,Valentinuzzi2011}. These data consist of $B$- and $V$-band observations obtained with the 2.5-meter Isaac Newton Telescope and the MPG/ESO 2.2-meter telescope \citep[see][for details regarding image and catalog access]{Moretti2014}. 

Compiling a sample of red and blue cluster spiral galaxies is not straightforward in terms of its impact on results. For example, \citet{Masters2010} has shown that the inclusion of edge-on galaxies tends to produce a redder spiral population due to extinction effects in the host galaxy \citep[see also][]{Kourkchi2019}. There is also the possibility of selecting S0-type galaxies if not careful, which can bias the conclusions of any study that is focused on the star formation activity of spiral galaxies \citep{Wolf2009}. Also, comparing spirals selected from low-density field environments to cluster spirals when examining the cluster galaxy population can introduce unknown selection bias, given that cluster spirals tend to be redder optically than field spirals \citep[e.g.,][]{Boselli2006,Cantale2016}. 

To minimize selection effects, we select red and blue cluster spirals relative to their position in the CMD. Galaxies included in our sample have been visually verified to be face-on or nearly face-on spirals with prominent spiral arms, eliminating the inclusion of non-spiral systems and edge-on galaxies. For each cluster, only galaxies with a published spectroscopic redshift (obtained from the NASA/IPAC Extragalactic Database) consistent with being within $\pm3\sigma_{\text{v}}$ \citep[as measured with respect to the cluster velocity dispersion;][]{Yahil1977} of the recessional velocity of the brightest cluster galaxy (BCG) are included in our sample. Adopted cluster velocity dispersions and references are tabulated in Table~\ref{tab:VelocityDispersions}. These selection criteria help to minimize contamination of our galaxy compilation from the inclusion of interlopers, S0 galaxies, and edge-on systems.

\begin{table*}
\scriptsize
\centering
\caption{Galaxy cluster velocity dispersions and reference sources.}
\begin{tabular}{|c|c|c|c|c|c|}\hline
Cluster & $\sigma_{v}$& Reference & Cluster & $\sigma_{v}$ & Reference\\
& (km s$^{-1}$) &  & & (km s$^{-1}$) &  \\ \hline
A21 & 621 & \citet{Struble1999} & A2124 & 801 & \citet{Moretti2015} \\
A84 & 769 & \citet{White1997} & A2147 & 821 & \citet{Barmby1998} \\
A85 & 1009 & \citet{Lauer2014} & A2149 & 353 & \citet{Moretti2015} \\
A98 & 812 & \citet{Rude2020} & A2152 & 456 & \citet{Lauer2014} \\
A98N & 690 & \citet{Rude2020} & A2169 & 509 & \citet{Moretti2015} \\
A119 & 901 & \citet{Lauer2014} & A2199 & 780 & \citet{Oegerle2001} \\
A133 & 790 & \citet{Lauer2014} & A2244 & 1240 & \citet{Struble1999} \\
A147 & 621 & \citet{Lauer2014} & A2247 & 353 & \citet{Lauer2014} \\
A151 & 795 & \citet{Lauer2014} & A2255 & 1266 & \citet{Struble1999} \\
A154 & 988 & \citet{Lauer2014} & A2256 & 1301 & \citet{Lauer2014} \\
A168 & 625 & \citet{Lauer2014} & A2382 & 900 & \citet{Lauer2014} \\
A193 & 776 & \citet{Lauer2014} & A2384 & 1051 & \citet{Pranger2014} \\
A260 & 754 & \citet{Lauer2014} & A2399 & 713 & \citet{Lauer2014} \\
A350 & 627 & \citet{Rude2020} & A2410 & 598 & \citet{Struble1999} \\
A351 & 510 & \citet{Popesso2007} & A2415 & 722 & \citet{Lauer2014} \\
A362 & 758 & \citet{Rude2020} & A2440 & 957 & \citet{Struble1999} \\
A376 & 830 & \citet{Lauer2014} & A2457 & 642 & \citet{Lauer2014} \\
A401 & 1161 & \citet{Lauer2014} & A2556 & 872 & \citet{White1997} \\
A496 & 737 & \citet{Lauer2014} & A2572 & 593 & \citet{Lauer2014} \\
A500 & 771 & \citet{Lauer2014} & A2589 & 872 & \citet{Lauer2014} \\
A514 & 1180 & \citet{Lauer2014} & A2593 & 644 & \citet{Lauer2014} \\
A548 & 795 & \citet{Lauer2014} & A2622 & 860 & \citet{Lauer2014} \\
A602 & 796 & \citet{Lauer2014} & A2626 & 648 & \citet{Lauer2014} \\
A634 & 331 & \citet{Lauer2014} & A2634 & 919 & \citet{Lauer2014} \\
A646 & 738 & \citet{Popesso2007} & A2657 & 807 & \citet{Lauer2014} \\
A655 & 736 & \citet{Popesso2007} & A2670 & 963 & \citet{Lauer2014} \\
A671 & 850 & \citet{Lauer2014} & A2688 & 643 & \citet{Rude2020} \\
A690 & 546 & \citet{Lauer2014} & A2717 & 568 & \citet{Lauer2014} \\
A754 & 995 & \citet{Lauer2014} & A2734 & 843 & \citet{Lauer2014} \\
A779 & 450 & \citet{Lauer2014} & A3128 & 892 & \citet{Lauer2014} \\
A780 & 871 & \citet{Lauer2014} & A3158 & 1095 & \citet{Lauer2014} \\
A795 & 778 & \citep{Rines2013} & A3266 & 1251 & \citet{Lauer2014} \\
A957 & 772 & \citet{Lauer2014} & A3376 & 855 & \citet{Lauer2014} \\
A970 & 841 & \citet{Lauer2014} & A3395 & 950 & \citet{Lauer2014} \\
A999 & 286 & \citet{Lauer2014} & A3490 & 996 & \citet{Lauer2014} \\
A1069 & 706 & \citet{Lauer2014} & A3497 & 761 & \citet{Lauer2014} \\
A1142 & 757 & \citet{Lauer2014} & A3528a & 726 & \citet{Moretti2015} \\
A1213 & 572 & \citet{Lauer2014} & A3528b & 961 & \citet{Lauer2014} \\
A1291 & 724 & \citet{Lauer2014} & A3530 & 716 & \citet{Lauer2014} \\
A1569 & 622 & \citet{Lauer2014} & A3532 & 734 & \citet{Lauer2014} \\
A1631a & 753 & \citet{Lauer2014} & A3556 & 657 & \citet{Lauer2014} \\
A1644 & 1016 & \citet{Lauer2014} & A3558 & 1002 & \citet{Lauer2014} \\
A1650 & 799 & \citet{Popesso2007} & A3560 & 261 & \citet{Lauer2014} \\
A1656 & 1035 & \citet{Lauer2014} & A3562 & 966 & \citet{Lauer2014} \\
A1668 & 649 & \citet{Moretti2015} & A3667 & 1028 & \citet{Lauer2014} \\
A1736 & 1127 & \citet{Lauer2014} & A3716 & 609 & \citet{Lauer2014} \\
A1775 & 568 & \citet{Lauer2014} & A3809 & 677 & \citet{Lauer2014} \\
A1795 & 861 & \citet{Lauer2014} & A3880 & 854 & \citet{Lauer2014} \\
A1831 & 1176 & \citet{Lauer2014} & A4059 & 830 & \citet{Lauer2014} \\
A1913 & 636 & \citet{Lauer2014} & IIZW108 & 513 & \citet{Moretti2015} \\
A1920 & 562 & \citet{Tovmassian2012} & MKW3s & 539 & \citet{Moretti2015} \\
A1940 & 785 & \citet{Struble1999} & RX0058 & 637 & \citet{Moretti2015} \\
A1983 & 541 & \citet{Lauer2014} & RX1022 & 577 & \citet{Moretti2015} \\
A1991 & 604 & \citet{Lauer2014} & RX1740 & 582 & \citet{Moretti2015} \\
A2022 & 607 & \citet{Lauer2014} & Z2844 & 536 & \citet{Moretti2015} \\
A2029 & 1222 & \citet{Lauer2014} & Z8338 & 712 & \citet{Moretti2015} \\
A2100 & 582 & \citet{Rude2020} & Z8852 & 765 & \citet{Moretti2015} \\
A2107 & 629 & \citet{Lauer2014} &   &&   \\ \hline
\end{tabular}
\label{tab:VelocityDispersions}
\end{table*}

Cluster galaxies are selected from the low-redshift cluster samples of \citet{LopezCruz2004}, \citet{Barkhouse2007}, \citet{Valentinuzzi2011}, and \citet{Rude2020}. All cluster catalogs have been uniformly analyzed to determine red-sequence slopes, y-intercepts, and dispersions following the procedure outlined in \citet{LopezCruz2004}. The faint-end cutoff is defined as the limiting magnitude where the average $2.5\sigma$ uncertainty in color, in the color-magnitude diagram, exceeds the $\pm 3\sigma$ dispersion of the red-sequence \citep{Barkhouse2007}. All galaxies fainter than this limit are not included in our final catalog. We define ``red'' galaxies as those that are within $\pm 3\sigma$ of the red-sequence, while ``blue'' galaxies are defined to be those $>3\sigma$ blueward of the red-sequence, where $\sigma$ is the color dispersion of the red-sequence. Our final sample consists of 324 red spirals and 273 blue spiral galaxies selected from 115 low-redshift galaxy clusters.

We stress that our galaxy selection is justified based on several important factors: a. Selecting only cluster spirals provides a less-biased sample for comparison since both red and blue cluster spirals, on the average, are H I deficient and have less star formation than field spirals \citep{Boselli2006}. b. Including only face-on spirals helps to minimize the impact that the host galaxy dust will have on optical colors of red galaxies compared to the inclusion of edge-on systems \citep{Wolf2009,Masters2010}. c. The selection of spirals with visually observable spiral arms imposes a stringent constraint on any process responsible for quenching star formation. The presence of spiral arms suggests that the quenching process did not violently disrupt the spiral structure, as would be expected if galaxy interactions or mergers were the primary mechanisms for transforming blue spirals into red galaxies. Furthermore, it indicates that the time elapsed since the quenching event has been brief \citep{Masters2010}. The primary objective is to obtain a well-defined sample of blue and red face-on cluster spirals to compare their properties, thereby facilitating a deeper understanding of galaxy evolution in cluster environments.

\section{Analysis and results}

We obtained multi-wavelength observations of our spiral sample by cross-correlating our red and blue spiral positions with {\it GALEX} \citep{Martin2005} $FUV$ and $NUV$ observations, {\it WISE} $W1$ ($3.4~\mu m$), $W2$ ($4.6~\mu m$), $W3$ ($12~\mu m$), and $W4$ ($22~\mu m$) photometry \citep{Wright2010}, and SDSS DR 17 $u$-, $g$-, $r$-, $i$-, and $z$-band imaging and spectroscopy \citep{Abdurrouf2022}. For cross-correlation, we used a positional matching radius of $3\arcsec$, with the nearest matched object retained if multiple objects were found within the $3\arcsec$ radius cone. A total of 140 blue spirals and 68 red spirals have photometry available for all bandpasses from {\it GALEX}, {\it WISE}, and SDSS. No correction for differences in the point spread function (PSF) were applied since galaxy sizes are $>3\times$ the largest PSF. This is consistent with the requirement that selected galaxies must be large enough in apparent size to display well-defined spiral arm structure. 

All magnitudes were corrected for Milky Way extinction. CFHT $u$ and $r$ mags have been photometrically calibrated based on SDSS deredden modelMags \citep{Rude2020}, which have been extinction corrected based on \citet{Schlafly2011} for $R_{v}=A_{v}/E(B-V)=3.1$ \citep{Cardelli1989,Fitzpatrick1999}. KPNO $B$ and $R_{c}$ magnitudes have been extinction corrected based on \citet{Burstein1982,Burstein1984}. WINGS $B$- and $V$-band data have been extinction corrected based on \citet{Schlafly2011} for $R_{v}=3.1$. {\it GALEX} $FUV$ and $NUV$, {\it WISE} $W1$, $W2$, $W3$, and $W4$ magnitudes were extinction corrected using \citet{Zhang2023} for $R_{v}=3.1$. We note that no attempt was made to correct for individual internal galaxy extinction. 

A k-correction was applied to each galaxy based on redshift and color (except for {\it WISE} data) as calculated from \citet{Chilingarian2010} and \citet{Chilingarian2012}. For {\it WISE} k-corrections we followed the procedure outlined in \citet{Jarrett2023} with data from \citet{Jarrett2011}.

\subsection{Optical properties}\label{sec:optproperties}
\begin{figure*}
\centering
\includegraphics[scale=1.0]{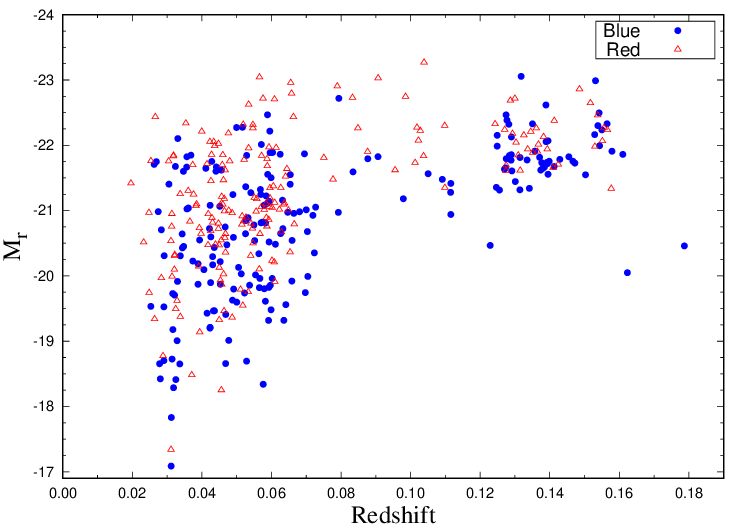}
\caption[Absolute Mag vs Redshift]{Redshift distribution of extinction- and k-corrected $M_{r}$. Red spirals are depicted by open red triangles, while blue spirals are represented by filled blue circles. Uncertainties in $M_{r}$ are smaller than plot symbols.}
\label{fig:Mr_redshift}
\end{figure*}

In Figure~\ref{fig:Mr_redshift} we show the redshift distribution of extinction- and k-corrected absolute magnitudes $M_{r}$ for our red and blue galaxy samples. Values of $M_{r}$ for red spirals range from $-17.3$ to $-23.3$ with a mean of $-21.2$, while blue galaxies have $-17.1 > M_{r} > -23.0$ and a mean of $-20.9$. Galaxies at lower redshifts are sampled to fainter $M_{r}$ compared to those at higher redshifts. The mean redshift of our blue galaxy compilation is $0.076$, which is similar to the mean redshift of the red galaxies ($z=0.067$).

\begin{figure}
\centering
\includegraphics[scale=1.0]{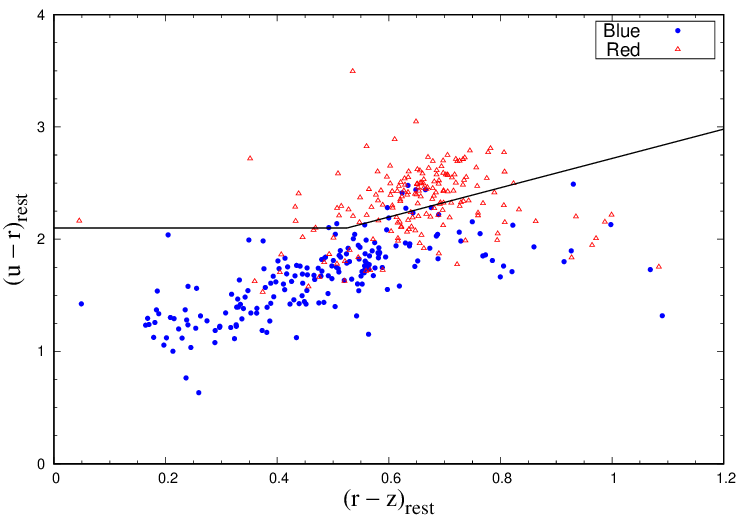}
\caption[SDSS Colors]{Rest-frame SDSS color-color distribution of red and blue spirals. Magnitudes are extinction-corrected (dereddened) model magnitudes. Black lines are from \citet{Chang2015} and separate star-forming and passive galaxies. Plot symbols are the same as those used in Figure~\ref{fig:Mr_redshift}. Color uncertainties are approximately the same size as plot symbols.}
\label{fig:SDSS-color-color}
\end{figure}

To look for differences in optical/near-optical colors, we show in Figure~\ref{fig:SDSS-color-color} the rest-frame (k-corrected) extinction-corrected SDSS $u-r$ vs. $r-z$ color-color diagram. The black lines depicted in Figure~\ref{fig:SDSS-color-color} are from \citet{Chang2015}, and represent the division between ``star-forming'' (below the lines) and ``passive'' (above the lines) galaxies. The median offset between the two distributions is 0.68 dex with an rms scatter of $\sim 0.3$ dex. For the blue spirals, $96\%$ are found in the star-forming region of the figure, while $59\%$ of red spirals are located in the passive region of the diagram. Using a two-sample, two-dimensional Kolmogorov-Smirnov (K-S) test, we find that the two samples are statistically distinct (see Table~\ref{tab:KSTests}).

\begin{figure}
\centering
\includegraphics[scale=1.4]{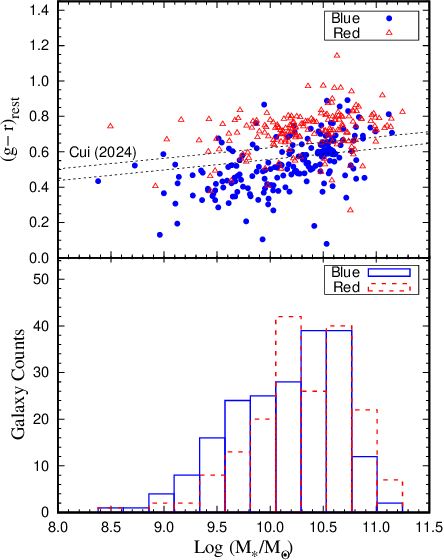}
\caption[]{{\it Top panel:} The extinction and k-corrected $g-r$ color-mass diagram. Dashed lines represent the green valley region as defined by \citet{Cui2024}. Plot symbols are the same as those used in Figure~\ref{fig:Mr_redshift}. Color uncertainties are approximately the size of plot symbols. {\it Bottom panel:}  Galaxy mass histogram based on extinction- and k-corrected absolute $r$-band values. Blue spirals (solid boxes) and red spirals (dashed boxes) show a similar distribution.}
\label{fig:Mass_Histogram_gmr}
\end{figure}

The top panel of Figure~\ref{fig:Mass_Histogram_gmr} depicts the extinction- and k-corrected $g-r$ color distribution with galaxy mass. Mass values are derived from \citet[][see their Figure 2]{Mahajan2018} based on extinction- and k-corrected $M_{r}$. The dashed lines enclose the green valley region as defined by \citet{Cui2024} for the MaNGA survey. We find that 16\% of blue spirals are found in the green valley strip, while 8.7\% of red spirals are located in this region. As evident from the top panel of Figure~\ref{fig:Mass_Histogram_gmr}, there is no clean separation of red and blue galaxies using $g-r$ color. For example, we find 79\% of red galaxies and 24\% of blue spirals have $g-r$ color redder than the green valley region defined by \citet{Cui2024}. In contrast, 12\% of the red galaxies and 60\% of the blue spirals have $g-r$ color blueward than the green valley as defined by \citet{Cui2024}. We note that \citet{Salim2014} has shown that $g-r$ color selection does a rather poor job of isolating green valley galaxies.

The bottom panel of Figure~\ref{fig:Mass_Histogram_gmr} gives the histogram mass distribution, showing that both blue and red spirals sample similar range in galaxy mass. Using $\log (M_{\ast}/M_{\odot}) > 10$ as a fiducial for massive galaxies, 46.8\% of massive galaxies are blue and 53.2\% are red.

\subsection{Ultraviolet properties}

\begin{figure}
\centering
\includegraphics[scale=1.1]{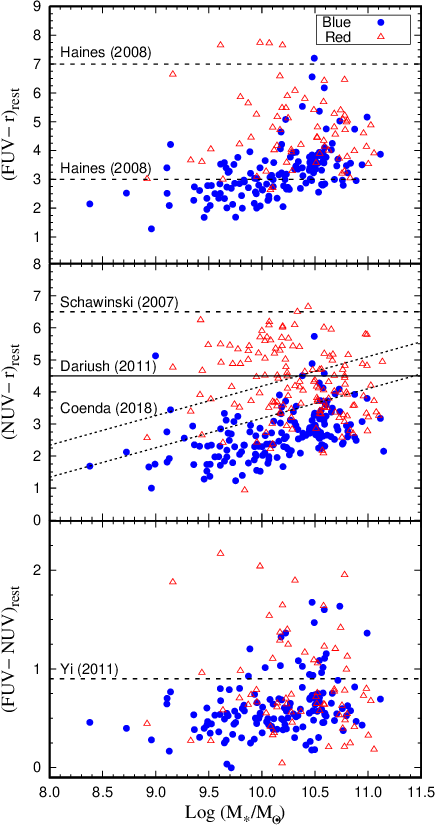}
\caption[]{{\it Top panel:}  $FUV-r$ color-mass diagram. Dashed lines are from \citet{Haines2008} and represent passive galaxies at $FUV-r=7$ and star-forming galaxies at $FUV-r=3$. {\it Middle panel:} $NUV-r$ color-mass diagram. The solid line depicts the separation between blue and red galaxies as defined by \citet{Dariush2011}. The area between the dotted lines depict the green valley region as defined by \citet{Coenda2018}. The dashed line at $NUV-r=6.5$ denotes the separation between passive and dusty red galaxies given by \citet{Schawinski2007}. {\it Bottom panel:} $FUV-NUV$ color-mass diagram. The dashed line depicts the upper limit for UV-upturn galaxies from \citet{Yi2011}. Plot symbols are the same as those used in Figure~\ref{fig:Mr_redshift}.}
\label{fig:UV_plots}
\end{figure}

Since ultraviolet-optical colors are sensitive to recent star formation \citep[e.g.,][]{Salim2007,Kennicutt2012}, we plot in Figure~\ref{fig:UV_plots} extinction-corrected and rest frame $GALEX$-SDSS $r$-band color-mass distributions for red and blue spirals. Galaxy masses are calculated in the same manner as described for Figure~\ref{fig:Mass_Histogram_gmr}. In the top panel of the figure, the dashed lines from \citet{Haines2008} represent red-sequence passive galaxies at $FUV-r=7$, while blue cloud star-forming galaxies are located at $FUV-r=3$. Our galaxies show a range in $FUV-r$ color for the red spirals, with 90\% of these galaxies having colors between $FUV-r=3$ and 7, with an average of 4.7 and a standard deviation of 1.2. Our blue galaxy sample has an average color of $FUV-r=3.2$ (standard deviation of 0.9), which is consistent with \citet{Haines2008}. We also note a trend that higher mass blue galaxies have a redder $FUV-r$ color.       
\begin{table*}
\centering
\caption{Kolmogorov-Smirnov test results.}
\begin{tabular}{|l|c|c|}\hline
Distribution & D Statistic & P-Value\\ \hline
($u-r$) vs. ($r-z$) & 0.72 & $2.71\times 10^{-3}$ \\
$D_{n}$(4000) Histogram & 0.70 & $6.32\times 10^{-3}$ \\
EW($H\delta$) vs. $D_{n}$(4000) & 0.67 & $8.43\times 10^{-3}$ \\
EW($H\delta$) vs. $H\alpha$ & 0.65 & $6.70\times 10^{-3}$ \\
$[$O~III] Histogram & 0.18 & $2.83\times 10^{-3}$ \\
$H\alpha$ Histogram & 0.45 & $2.21\times 10^{-3}$ \\
$[$O III] vs. $H\alpha$ & 0.43 & $4.74\times 10^{-3}$ \\
sSFR vs. Galaxy Mass & 0.50 & $4.67\times 10^{-3}$ \\
sSFR Histogram & 0.67 & $4.89\times 10^{-3}$ \\
sSFR vs. ($M_{g}-M_{r}$) & 0.66 & $7.93\times 10^{-3}$ \\
EW($H\alpha$) vs. $D_{n}$(4000) & 0.71 & $1.41\times 10^{-3}$ \\
$(\Delta v/\sigma_{v})\times (r/r_{200})$ vs. $r/r_{200}$ & 0.20 & $5.29\times 10^{-4}$ \\ \hline
\end{tabular}
\label{tab:KSTests}
\end{table*}

For the middle panel, the dashed line at $NUV-r=6.5$ was used by \citet{Schawinski2007} to separate dusty red galaxies from non-dusty systems. Only two red galaxies (1.4\%) have $NUV-r >6.5$, indicating that red spirals are not red due to an overabundance of dust. The solid line at $NUV-r=4.5$ was used by \citet{Dariush2011} to separate passive and star-forming galaxies. We find 3\% (42\%) of our blue (red) galaxies have $NUV-r>4.5$. The dotted diagonal lines outline the green valley region as defined by \citet{Coenda2018}. Galaxies in this area are possibly transitioning from blue star-forming systems to red passive galaxies. For our sample, we find 12\% of blue galaxies and 31\% of red spirals are found in the green valley region. \citet{Haines2008} separate passive and star-forming galaxies using $NUV-r=4$, noting that very few star-forming galaxies have $NUV-r>4$. Using this dividing color, we find that 95\% of blue spirals have $NUV-r<4$ and 55\% red spirals have $NUV-r>4$. Red galaxies have a $NUV-r$ dispersion of 0.89, while the blue sample has a smaller dispersion of 0.56. For the blue spirals, we also observe a trend that $NUV-r$ color becomes redder with increasing galaxy mass (recall that this is also observed for $FUV-r$ depicted in the top panel of Figure~\ref{fig:UV_plots}).         

In the bottom panel of Figure~\ref{fig:UV_plots} we show the mass distribution of galaxy $FUV-NUV$ color. Red and blue spirals are not well separated by $FUV-NUV$ color. Using $FUV-NUV=0.9$ as an upper limit for UV-upturn galaxies \citep[dashed line;][]{Yi2011}, we find that 86\% of blue and 66\% of red spiral galaxies have a $FUV-NUV$ color below this upper limit. Therefore, 21\% of the combined red+blue spirals have $FUV-NUV>0.9$ and are found redward of the UV-upturn upper limit \citep{Boissier2018}.

\subsection{WISE infrared measurements}

\begin{figure}
\centering
\includegraphics[scale=1.0]{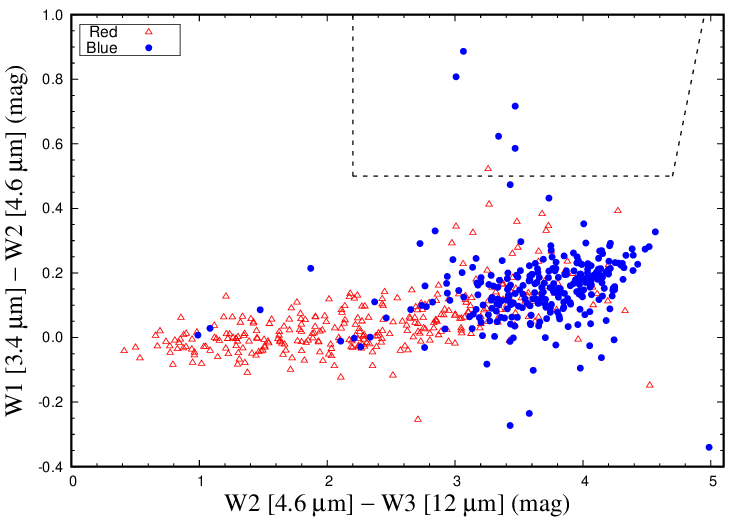}
\caption[]{{\it WISE} color-color diagram for red and blue spiral galaxies. Blue galaxies (filled circles) dominate the $[4.6]-[12]>3$ region, while red galaxies (open triangles) extend over a much larger range in $[4.6]-[12]$ color. The area enclosed by the dashed lines is the AGN region defined by \citet{Jarrett2017}.}
\label{fig:WISE}
\end{figure}

Cross-matching galaxy positions with {\it WISE} data allow us to use mid-infrared observations to discriminate between passive and star-forming systems. In Figure~\ref{fig:WISE} we plot the extinction- and k-corrected $W1[3.4~\mu m]-W2[4.6~\mu m]$ vs. $W2[4.6~\mu m]-W3[12~\mu m]$ color-color galaxy distribution (Vega magnitudes). Blue spirals are found mainly with  $W2-W3>3$, while red spirals extend over a larger range from approximately $0.4 <W2-W3<4.5$. Comparing our Figure~\ref{fig:WISE} with Figure 12 from \citet{Wright2010}, we find that our red spirals cover the region of the color-color plot that encompasses elliptical/spiral galaxies, while blue spirals stretch mainly from spiral- to starburst-type systems. 

\citet{Cui2024} separated massive red spirals into quenched and star-forming dusty systems using $W2-W3=2.5$ as the dividing color. For our red galaxy sample, we find 45\% have $W2-W3>2.5$, indicating that these galaxies are red in the optical due to dust. Since most of our blue spirals have $W2-W3>3$ (91\%), adapting $W2-W3=3$ as a conservative color cutoff between passive and star-forming dusty red galaxies, we find that 29\% of red spirals have a $W2-W3$ color $>3$. This is similar to the result from \citet{Cui2024}, where they find 23\% of their red sequence spirals have $W2-W3>2.5$ and are thus classed as dusty spirals. The finding that 91\% of blue spirals and 29\% of red spirals have $W2-W3>3$ implies that blue spirals are dusty on average compared to red-sequence spiral galaxies. This result has also been found by other studies \citep[e.g.,][]{Mahajan2020}. 

Using $W2-W3$ color to separate ellipticals/intermediate disks/star-forming disk galaxies \citep{Jarrett2017}, we find that 37\% of red spirals have $W2-W3<2$ (ellipticals), 52\% have $2<W2-W3<3.5$ (intermediate disks), and 11\% have $W2-W3>3.5$ (star-forming disks). Using the same color cuts for blue spirals, 1.5\% are ellipticals, 33\% are intermediate disks, and 65.5\% are star-forming disk systems. 

The region enclosed by the dashed lines in Figure~\ref{fig:WISE} outlines the area occupied predominantly by AGN according to \citet{Blecha2018} and \citet{RamosPadilla2020}. AGN feedback has been associated with the quenching of star formation in spiral galaxies, ultimately making them red compared to star-forming spirals \citep[negative feedback; e.g.,][]{Ryzhov2025}. It has also been found that AGN feedback can compress star-forming gas and enhance star formation \citep[positive feedback; e.g.,][]{Zhuang2020}. For our galaxy sample, we find five blue spirals (1.8\%) and one red spiral (0.3\%) in the AGN region of the {\it WISE} color-color plot. Detailed examination of the detection and role of AGN feedback in our red and blue spirals can be found in \citet{Akter2026}.  

\subsection{SDSS spectroscopic analysis}

Spectroscopic analysis was conducted using SDSS data from the {\tt galSpecLine} database table based on the reanalysis of emission line measurements from the MPA-JHU compilation of \citet{Brinchmann2004} and \citet{Tremonti2004}. Galaxy nebular emission line measurements have been corrected for Galactic extinction based on \citet{O'Donnell1994} and \citet{Schlegel1998}. In addition, data from the {\tt emissionLinesPort} and {\tt galSpecIndx} SDSS tables based on \citet{Thomas2013} were used. As part of the {\tt galSpecIndx} catalog, we used the D\textsubscript{n}(4000) spectral index \citep{Balogh1999} as a proxy for stellar age and hence star formation. This spectral index was calculated by taking the ratio of the red continuum (as measured between 4000 and 4100 {\AA}) to the blue continuum (3850 to 3950 {\AA}). In general, a larger value of D\textsubscript{n}(4000) implies a greater stellar population age and a lower star formation rate. 

Since SDSS fibers probe a circular aperture of 3\arcsec in diameter, D\textsubscript{n}(4000) provides information related to central galaxy star formation. Given the redshift range of our galaxy sample, the fiber physical coverage ranges from 0.8 kpc to 9.0 kpc in diameter. Although SDSS spectroscopic data only includes the central region of galaxies, the main goal of this research is to compare and contrast red and blue spirals, looking for any {\it relative differences} rather than obtaining absolute measurements for each galaxy as a whole. For all spectroscopic analysis, we only include galaxies that have measured spectral lines with $S/N>3$. In total, SDSS spectroscopic measurements are available for 154 blue spirals and 122 red spirals from our galaxy sample.

\begin{figure}
\centering
\includegraphics[scale=1.0]{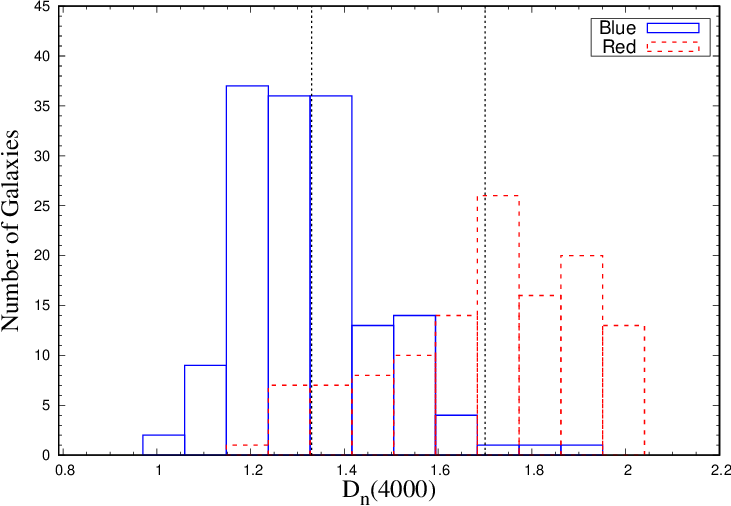}
\caption[]{Histogram distribution of D\textsubscript{n}(4000) for blue (solid boxes) and red spiral galaxies (dashed boxes). On average, blue spirals have a lower stellar age and hence more recent star formation compared to red spiral systems. The two dashed vertical lines depict the average value of D\textsubscript{n}(4000) for the two populations (D\textsubscript{n}(4000)$\,=1.33\pm 0.14$ for blue spirals and D\textsubscript{n}(4000)$\,=1.70\pm 0.21$ for red galaxies).}
\label{fig:Dn4000_Histogram}
\end{figure}

In Figure~\ref{fig:Dn4000_Histogram}, we plot the histogram distribution of D\textsubscript{n}(4000) for blue and red spirals. The average D\textsubscript{n}(4000) value for blue galaxies is $1.33\pm 0.14$ and $1.70\pm 0.21$ for red spirals (depicted by the dashed vertical lines in Figure~\ref{fig:Dn4000_Histogram}). These values are similar to the results of \citet[][see their Figure 6]{Masters2010}. A one-sample K-S test indicates that blue and red spirals are unlikely drawn from the same parent population (see Table~\ref{tab:KSTests}). The larger average value of D\textsubscript{n}(4000) for red spirals is consistent with these galaxies being mainly passive systems with very little recent star formation. Using a value of D\textsubscript{n}(4000)$\,<1.4$ as an indicator of galaxies with average stellar age < 1 Gyr \citep{Kauffmann2003}, we find 75\% of blue spirals and only 11\% of red spirals have D\textsubscript{n}(4000)$\,<1.4$.

\begin{figure}
\centering
\includegraphics[scale=0.9]{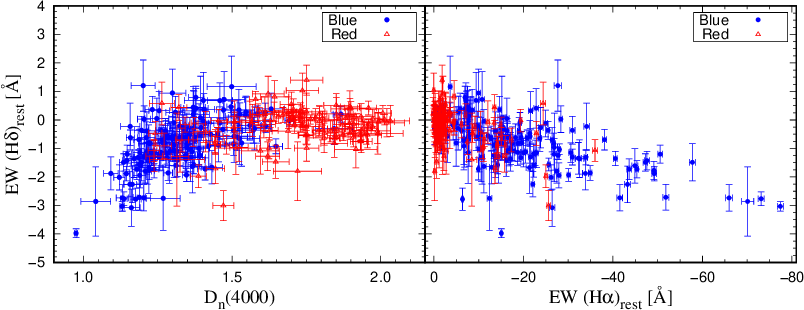}
\caption[]{{\it Left panel:} Rest frame H$\delta$ equivalent width versus D\textsubscript{n}(4000) spectral index. {\it Right panel:} Rest frame equivalent width of H$\delta$ versus rest frame H$\alpha$. Plot symbols in both panels are the same as that used in Figure~\ref{fig:Mr_redshift}.}
\label{fig:Hdelta_D4000}
\end{figure}

We plot in Figure~\ref{fig:Hdelta_D4000} the equivalent width (EW) of H$\delta$ (rest frame) versus D\textsubscript{n}(4000) (left panel) and versus EW of H$\alpha$ (rest frame; right panel). The adopted EW sign convention is absorption lines have a positive EW and emission lines have a negative EW. H$\delta$, along with D\textsubscript{n}(4000), is also a good indicator of the $4000\,\text{\AA}$ break and thus a proxy for stellar population age \citep{Kauffmann2003}. In particular, H$\delta$ is most prominent 0.1-1 Gyr after a starburst phase \citep{Shimakawa2022}. 

In the left panel of Figure~\ref{fig:Hdelta_D4000}, we find that blue spirals mostly have small values of D\textsubscript{n}(4000), and have H$\delta$ in emission (81\%), supporting a mainly star-forming stellar population. In contrast, red spirals on average are found with larger values of D\textsubscript{n}(4000) and less negative EW(H$\delta$) compared to blue spirals (63\% of red spirals have EW(H$\delta)<0\,\text{\AA}$). Using EW(H$\delta)=-1\,\text{\AA}$ as a dividing line, we find that 42\% of blue spirals have EW(H$\delta)<-1\,\text{\AA}$ and only 9.8\% of red spirals have EW(H$\delta)<-1\,\text{\AA}$. A two-sample, two-dimensional K-S test for H$\delta$ vs. D\textsubscript{n}(4000) shows that statistically the red and blue spirals are not from the same parent population (see Table~\ref{tab:KSTests}).

In the right panel of Figure~\ref{fig:Hdelta_D4000}, we plot the EW of H$\delta$ versus EW of H$\alpha$. H$\alpha$ emission is a well-known diagnostic indicator of ongoing star formation \citep[e.g.,][]{Kennicutt1998,Kennicutt2012}. Using EW(H$\alpha)=-2\,\text{\AA}$ as a dividing line between passive and star-forming galaxies \citep{Haines2007}, we find 95\% blue spirals are star-forming and 54\% of red spirals are passive systems. A two-sample, 2D K-S test for H$\delta$ vs. H$\alpha$ indicates that the red and blue spirals are most likely not from the same parent distribution (see Table~\ref{tab:KSTests}).

\subsection{Emission line properties}

\begin{figure}
\centering
\includegraphics[scale=1.0]{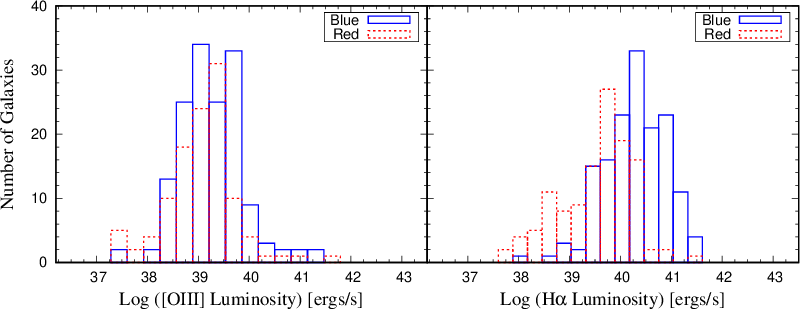}
\caption[]{{\it Left panel:} Extinction-corrected histogram distribution of log~L~[O III]. A significant overlap is visible between the red and blue galaxy populations. {\it Right panel:} Extinction-corrected histogram distribution of log~L~(H$\alpha$). The average values of log~L~(H$\alpha$) for the red and blue spirals are separated by 0.8 dex.}
\label{fig:Luminosity_Histogram}
\end{figure}

The luminosity of [O III] at $5007\,\text{\AA}$ is a tracer of both AGN activity and star formation of massive O-type stars \citep{Kauffmann2003}. In this study, we used the luminosity of [O III] to look for differences between red and blue spirals. In Figure~\ref{fig:Luminosity_Histogram} (left panel), we show the histogram distribution of the luminosity of [O III] $5007\,\text{\AA}$ for the two spiral compilations. There is a large overlap between the blue and red spiral populations, which may be influenced by the presence of AGN activity in both galaxy samples since the [O III] flux values are measured from SDSS fibers centered on each galaxy (additional details available in Akter et al., 2026). Using a dividing line of log~L~[O III] = 39.0, we find that 66\% of blue spirals and 61\% of red galaxies have log~L~[O III] > 39.0. A one-sample K-S test implies that the red and blue galaxies are statistically not part of the same parent population (see Table~\ref{tab:KSTests}).

In the right panel of Figure~\ref{fig:Luminosity_Histogram}, we show the histogram distribution of the luminosity of H$\alpha$ for our galaxy samples. Since H$\alpha$ emission is an indicator of recent star formation, we expect different mean values for H$\alpha$ luminosity if red spirals are mainly passive systems and blue spirals are active star-forming galaxies. For the red spirals, we find an average log~L~(H$\alpha)= 39.5$, while for the blue galaxies we have log~L~(H$\alpha)= 40.3$. Using a dividing line of log~L~(H$\alpha)= 40.0$, we find 69\% of blue spirals have log~L~(H$\alpha)>40.0$, while only 25\% of red spirals have log~L~(H$\alpha)>40.0$. This is consistent with red spirals being associated with passive galaxies and blue spirals with star-forming systems. A one-sample K-S test indicates that the galaxies are most likely not selected from the same parent population (see Table~\ref{tab:KSTests}).

\begin{figure}
\centering
\includegraphics[scale=1.0]{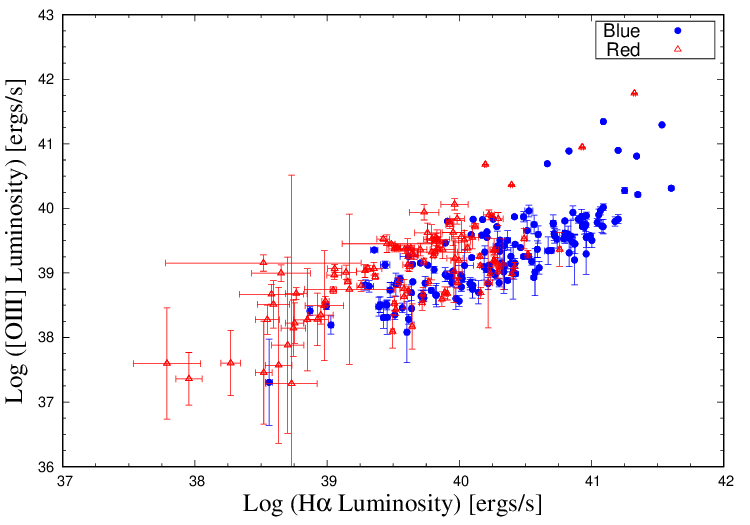}
\caption[]{Extinction-corrected [O III] luminosity versus H$\alpha$ luminosity. Red spirals, on average, have smaller values of both [O III] and H$\alpha$ luminosities compared to blue galaxies. Plot symbols are the same as those used in Figure~\ref{fig:Mr_redshift}.}
\label{fig:Luminosity_OIII_Halpha}
\end{figure}

To more clearly distinguish passive and star-forming galaxies, we plot the luminosity of [O III] versus L~(H$\alpha$) in Figure~\ref{fig:Luminosity_OIII_Halpha}. The majority of red spirals have a lower value of both [O III] and H$\alpha$ luminosity compared to blue spirals. This result is similar to that from \citet{Dhiwar2023}, where they plotted red, green valley, and blue elliptical galaxies (see their Figure 3; right panel). A two-sample, 2D K-S test shows that the red and blue galaxies are statistically not from the same source distribution (see Table~\ref{tab:KSTests}).

\subsection{Star formation rate}

\begin{figure}
\centering
\includegraphics[scale=1.0]{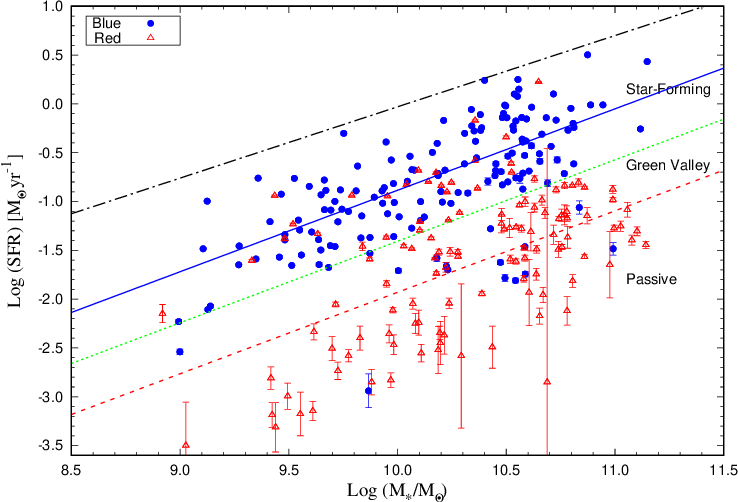}
\caption[]{Star formation rate versus galaxy mass. Blue galaxies are found mainly in the star-forming region, with red galaxies dominating the green valley and passive galaxy zones. The blue solid line is a fit to the blue spirals and denotes the star-forming region. The green dotted line is $1\sigma$ below the best-fit blue spiral line and represents the upper limit of the green valley. The dashed red line is $2\sigma$ below the best-fit blue line and defined as the upper envelope of the passive galaxy zone. The black dash-dotted line near the top of the figure depicts the best-fit straight line to the galaxy main sequence of the MaNGA data from \citet{Belfiore2018}. Plot symbols are the same as those used in Figure~\ref{fig:Mr_redshift}.}
\label{fig:SFR_Mass}
\end{figure}

Several methods and techniques have been used to estimate the star formation rate (SFR) of galaxies \citep[e.g.,][]{Kennicutt1998,Kennicutt2012}. For this study, we use the relation from \citet{Bell2001} and \citet{Dhiwar2023} based on H$\alpha$ luminosity, i.e., SFR~$(M_{\odot}\,yr^{-1})=7.93\times 10^{-42}~\text{L}\,(H\alpha)$, where L~(H$\alpha$) is in erg~s$^{-1}$. We plot the star formation rate versus galaxy mass in Figure~\ref{fig:SFR_Mass}, where galaxy mass is estimated based on extinction- and k-corrected $M_{r}$ absolute magnitudes \citep[e.g., log~M $=0.45+(-0.464\,M_{r})$;][]{Mahajan2018}. Many studies have used the SFR-$M_{*}$ plot to investigate the evolution of stars in galaxies, including understanding the increase in SFR with respect to galaxy mass \citep[so-called ``main sequence''; e.g.,][]{Brinchmann2004,Noeske2007,Renzini2015,Popesso2018,Bluck2020,Boselli2022}. 

For the data presented in Figure~\ref{fig:SFR_Mass}, a straight line was fit to the blue galaxy distribution using linear least-squares applied to nine equal mass bins using a biweight estimator \citep{Belfiore2018}. The resultant fit, depicted by the solid blue diagonal line, is given by  log~(SFR) = $(0.83\pm 0.05)\times \text{log} (M_{*})-9.23\pm 0.52$. Using the $1\sigma$ uncertainty in the y-intercept to define regions within the SFR-$M_{*}$ plot, we designate the area within $\pm 1\sigma$ of the straight line fit to the blue galaxies as the ``star-forming'' zone. The area $\Delta \text{SFR}=1\sigma$ below this region is defined as the ``green valley'' (i.e., the area between the green dotted line and the red dashed line), and the area below the green valley (below the red dashed line) is denoted as the ``passive'' region. 

In the star-forming region, we find that 137 (89.5\%) blue spirals and 29 (24\%) red spirals are located in this area. For the green valley transition region, we find 9 (5.9\%) blue spirals and 38 (31.4\%) red spirals. In the passive area, 7 (4.6\%) blue spirals and 54 (44.6\%) red spirals occupy this region. In summary, 89.5\% of the blue spirals are located in the star-forming region of the SFR-$M_{*}$ diagram, while 76\% of the red spirals are found in the green valley+passive regions. A two-sample, 2D K-S test indicates that the blue and red spirals are not from the same parent population (see Table~\ref{tab:KSTests}).

To compare our SFR-$M_{*}$ plot with published studies, we include in Figure~\ref{fig:SFR_Mass} a fit to star-forming galaxies (main sequence) from the MaNGA survey \citep[e.g.,][]{Bundy2015} as given by \citet[][upper black dash-dotted line]{Belfiore2018}. In comparing our results with \citet{Belfiore2018}, we find that our blue galaxies have a systematically lower SFR by a factor of $\sim 10$ (1 dex) over all sampled galaxy masses. Differences in the computation of star formation rates and the selection of galaxies may explain some of this difference. For example, the MaNGA survey uses integral field spectroscopy (IFS) to estimate the SFR, while our results are based on SDSS $H\alpha$ fiber measurements from the inner regions of galaxies. In addition, the MaNGA galaxy sample comprises objects selected from various environments, from low-density field regions to dense galaxy clusters. Recall that our sample only includes galaxies selected from low-redshift galaxy clusters.  

Galaxy clusters are well-known to contain spirals with a systematically lower SFR compared to similar mass spirals residing in low-density regions \citep[e.g.,][]{Boselli2006}. Clusters also contain spirals that are deficient in H I compared to field spirals \citep[cf.,][]{Denes2016,Boselli2022}. The suppression of star formation in cluster spirals may be due to galaxy interactions or feedback from AGN. The removal or heating of star-forming gas will suppress star formation and depress the SFR \citep[e.g.,][]{Man2018,Kalinova2021,Xu2022}. In \citet{Akter2026}, we address the impact of AGN feedback on star formation in our sample of blue and red spirals. Although the absolute calibration of the SFR for our spirals is not available, we stress that the primary goal of our study is to contrast and compare {\it relative differences} between red and blue cluster spirals.

In Figure~\ref{fig:SFR_Mass}, the well-known trend of increasing SFR with increasing galaxy mass is evident as depicted by the positive slope. To normalize the SFR with respect to galaxy mass, we divide the SFR by mass to construct the specific star formation rate (sSFR). Galaxy masses are estimated, as described previously (see Section~\ref{sec:optproperties}), using Figure 2 from \citet{Mahajan2018}. 

\begin{figure}
\centering
\includegraphics[scale=1.0]{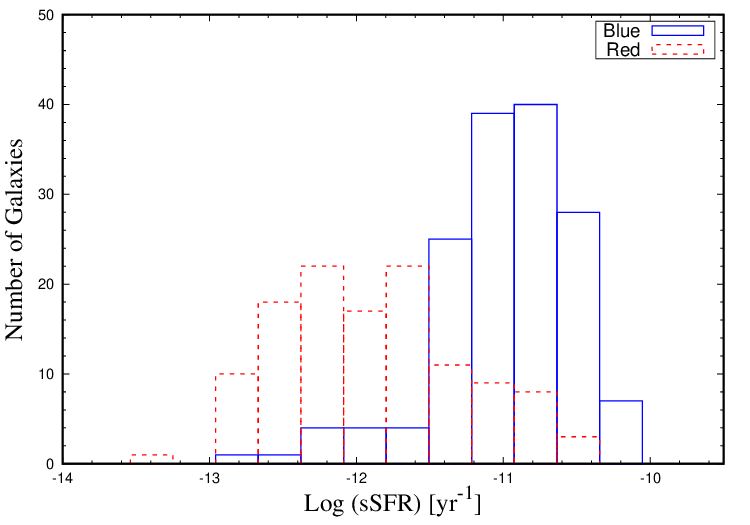}
\caption[]{Histogram distribution of the specific star formation rate for the blue and red spirals. As illustrated from the figure, blue spirals on average have a higher sSFR than red galaxies.}
\label{fig:sSFR_Histogram}
\end{figure}

In Figure~\ref{fig:sSFR_Histogram}, we present the histogram distribution of sSFR. From the figure we see a clear separation of the blue and red spirals, with blue spirals having a larger sSFR than the red galaxies. For the blue spirals, we find a median value of log~(sSFR)$\,=-10.9$, while for the red galaxies, we have log~(sSFR)$\,=-11.9$.

Fitting a Gaussian function to both the red and blue histograms, we find a local minimum between the two distributions at log~(sSFR)$\,=-11.47$. Using this dividing line, we identify that 90\% of the blue spirals have log~(sSFR)$\,>-11.47$ and only 25\% of red spirals have log~(sSFR)$\,>-11.47$. A one-sample K-S test shows that the red and blue spirals are statistically not from the same source population (see Table~\ref{tab:KSTests}).

\begin{figure}
\centering
\includegraphics[scale=1.0]{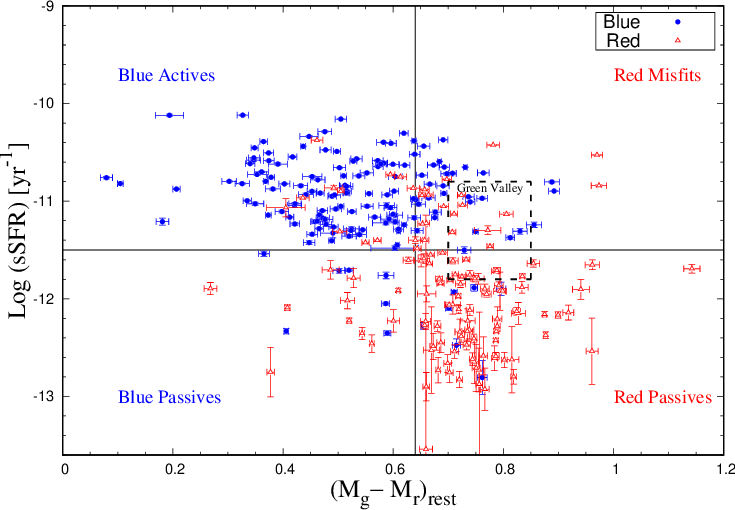}
\caption[]{Specific star formation rate versus $g-r$ absolute magnitude color. Magnitudes have been extinction- and k-corrected. Plot symbols are the same as those used in Figure~\ref{fig:Mr_redshift}.}
\label{fig:Mg_MrvssSFR}
\end{figure}

In Figure~\ref{fig:Mg_MrvssSFR}, we plot the specific star formation rate versus extinction- and k-corrected $M_{g}-M_{r}$ color. Using the same procedure as \citet{Evans2018}, we divide our plot into four quadrants based on $M_{g}-M_{r}=0.64$ and log~(sSFR)$ =-11.47$, which are determined from a fit to the histogram distribution in Figure~\ref{fig:sSFR_Histogram} and the $M_{g}-M_{r}$ color used by Evans et al. (see their Figure 2).

Using the same nomenclature as \citet{Evans2018}, in Figure~\ref{fig:Mg_MrvssSFR} the red star-forming galaxies ({\it red misfits}) are found in the upper-right region, red quiescent spirals ({\it red passives}) are located in the bottom-right area, blue quiescent spirals ({\it blue passives}) are in the bottom-left zone, and blue star-forming spirals ({\it blue actives}) are in the upper-left region. We also indicate the green valley region, depicted by the dashed line box, as defined by \citet{Salim2014} with $0.70\leq g-r\leq 0.85$ and $-11.8\leq~\mbox{log (sSFR)}\leq -10.8$. For the red misfits, we find 16\% of the red spirals and 29\% of the blue galaxies occupy this area. In the red passive region, 64\% of red spirals and 5.2\% of blue systems are located in this quadrant. For the blue ``passives'' zone, 9.9\% of red spirals and 4.6\% of the blue spirals are found in this region. For the blue ``actives'' quadrant, we find 9.1\% are red spirals and 71\% are blue spirals. For galaxies in the green valley region, we find $71\%$ are red spirals and $29\%$ are blue spirals.

A two-sample, 2D K-S test indicates that the blue and red spirals are most likely not selected from the same parent distribution (see Table~\ref{tab:KSTests}).

\begin{figure}
\centering
\includegraphics[scale=1.0]{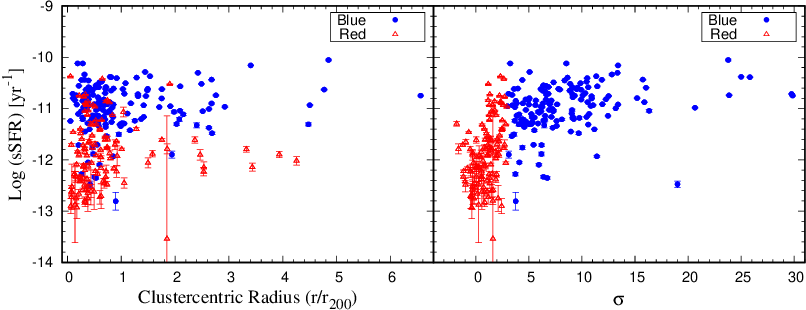}
\caption[]{{\it Left panel:} Specific star formation rate versus clustercentric radius ($r/r_{200}$). {\it Right panel:} Specific star formation rate versus distance from the host cluster red-sequence as measured from the red-sequence dispersion ($\sigma$) in color space. Plot symbols are the same as those used in Figure~\ref{fig:Mr_redshift}.}
\label{fig:sSFR_r200Sigma}
\end{figure}

The impact of the local environment on star formation has been studied for many years \citep[e.g.,][]{Balogh1998,Gomez2003}. We plot in Figure~\ref{fig:sSFR_r200Sigma} (left panel) the specific star formation rate versus clustercentric radius normalized to $r_{200}$ (i.e., $r/r_{200}$). The $r_{200}$ radius is the radius of a sphere within which the density is 200 times the critical density of the universe, and is used to normalize galaxy cluster radii \citep[cf.][]{Barkhouse2007}. The left panel of Figure~\ref{fig:sSFR_r200Sigma} indicates that the average of the sSFR is not correlated with clustercentric radius ($r/r_{200}$) for either red or blue spiral galaxies. The average value for the blue spirals is $\text{log (sSFR)}=-10.79$, and for the red spirals it is $\text{log (sSFR)}=-11.13$. We also find that the range in values of log~(sSFR) for both blue and red spirals at a small clustercentric radius is greater than for larger values of ($r/r_{200}$). Using the Pearson statistical measure, we find for the red spirals $r=0.0327$ and $p=0.7211$, and for the blue spirals $r=0.1383$ and  $p=0.0882$. For the Kendall $\tau$ statistic, we have for the red spirals $\tau=0.1307$ and $p=0.0335$, and for the blue spirals we find $\tau=0.0206$ and $p=0.7050$. These results indicate that there is a weak correlation between sSFR and clustercentric radius for both the red and blue spirals. 

In the right panel of Figure~\ref{fig:sSFR_r200Sigma}, we depict sSFR versus $\sigma$, where $\sigma$ is measured from the dispersion of the host cluster red-sequence (see Section 3.2). Recall that spiral galaxies $>3\sigma$ blueward of the red-sequence are classified as blue spirals. From the figure, we find that sSFR is not correlated with $\sigma$, indicating that blue spirals that are farthest from the red-sequence in color space have on average the same range in sSFR as blue spirals that are closer to the red-sequence. 

\subsection{Aging diagram}

\begin{figure}
\centering
\includegraphics[scale=1.0]{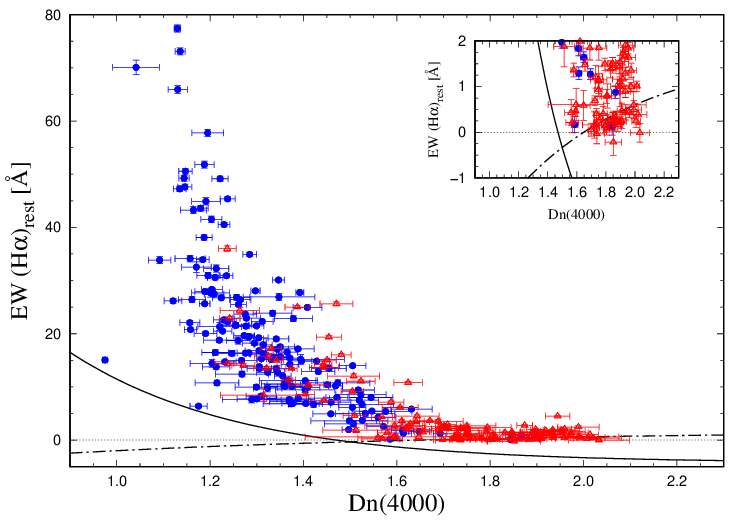}
\caption[]{Aging diagram using EW(H$\alpha$) versus D\textsubscript{n}(4000) for the red and blue spirals. The solid and dot-dashed curves are defined by \citet{Corcho-Caballero2023b} and separate regions containing ``aging,'' ``undetermined,'' and ``retired'' galaxies. The inset depicts a zoomed view of the y-axis to better display the retired galaxy region. Plot symbols are the same as those used in Figure~\ref{fig:Mr_redshift}.} 
\label{fig:AgingPlot}
\end{figure}

Aging diagrams have been used to aid the tracing of star formation history \citep[e.g.,][]{Corcho-Caballero2023a,Corcho-Caballero2023b,Privatus2025}. These diagrams generally use EW(H$\alpha$) versus D\textsubscript{n}(4000) to help compare star formation over the past 20 million years \citep[sensitivity of H$\alpha$ emission;][]{Corcho-Caballero2023a} and over a longer timescale of up to approximately 1 Gyr (D\textsubscript{n}(4000) diagnostic). This separation of star formation timescales allows the discrimination between current star-forming galaxies and passive systems \citep{Privatus2025}.

In Figure~\ref{fig:AgingPlot}, we plot the aging diagram for our red and blue galaxies using rest-frame EW(H$\alpha$) and D\textsubscript{n}(4000). The solid curve is described by equation 3 from \citet{Corcho-Caballero2023b}, EW(H$\alpha)=250.0\times 10^{-1.2~D_{n}(4000)}-4.3$, and the dot-dashed curve by equation 4, EW(H$\alpha)=-12.0\times 10^{-0.5~D_{n}(4000)}+1.8$ (note that equivalent widths are given as positive values to be consistent with published literature). According to Corcho-Caballero et al., galaxies located above the two curves are classified as ``aging'' galaxies. For our sample, we find 153 (99.4\%) of blue spirals and 101 (82.8\%) of red spirals are located in the aging region. This area contains galaxies that are presumably undergoing secular evolution with no abrupt changes to their star formation history \citep{Corcho-Caballero2023a}. 

Galaxies found below the solid curve and above the dot-dashed curve are so-called ``undetermined'' systems. There is a dearth of blue and red spiral galaxies in this part of the aging diagram. There is also a lack of galaxies in the area below the two curves, which is known as the ``quenched'' region. 

Galaxies located above the solid curve and below the dot-dashed curve (right half of Figure~\ref{fig:AgingPlot}) are classified as ``retired'' systems. The inset in Figure~\ref{fig:AgingPlot} shows a magnified view of the y-axis to better illustrate the retired galaxy section. For this area, we find 1 (0.6\%) blue spiral and 21 (17.2\%) red spirals. Retired galaxies are those that have experienced a truncation of star formation during the past approximately 1 Gyr \citep{Corcho-Caballero2023a}. This region of the aging diagram is clearly dominated by red spirals.

Finally, a two-sample, 2D K-S test yields that the blue and red spirals are statistically not selected from the same parent distribution in the aging diagram (see Table~\ref{tab:KSTests}).

\subsection{Phase-space diagram} 

\begin{figure}
\centering
\includegraphics[scale=1.0]{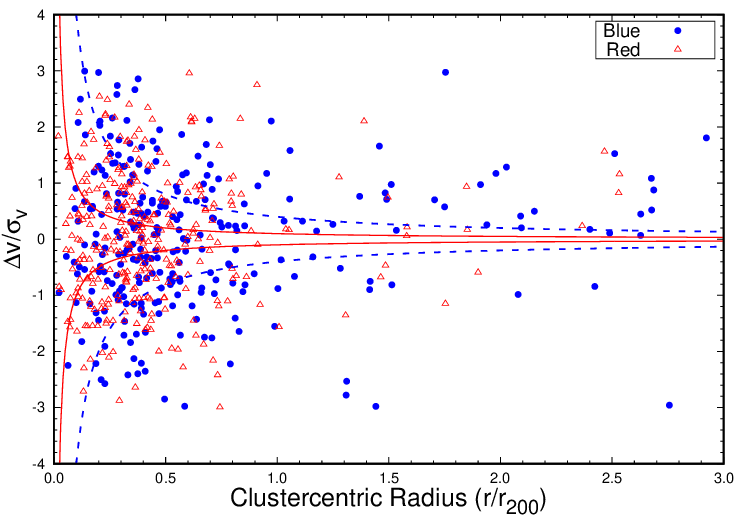}
\caption[]{Projected phase space diagram for red and blue spirals. The red curves depict the $(\Delta v/\sigma_{v})\times (r/r_{200})=0.1$ caustic profile, while 
the blue dashed curves represent the $(\Delta v/\sigma_{v})\times (r/r_{200})=0.4$ caustic profile. Plot symbols are the same as those used in Figure~\ref{fig:Mr_redshift}.}
\label{fig:PhaseSpace}
\end{figure}

The projected phase-space diagram \citep{Noble2013,Boselli2022}, a plot of line-of-sight galaxy velocity normalized with respect to the average cluster velocity dispersion ($\Delta v/\sigma_{v}$) versus projected clustercentric radius ($r/r_{200}$), has been used to help statistically discriminate galaxies that are recent infalls from others that have become virialized in the central regions of clusters or are backsplash galaxies \citep[e.g.,][]{Mahajan2011,Rhee2017,Shimakawa2022}. In Figure~\ref{fig:PhaseSpace} we plot the projected phase-space diagram for our red and blue galaxy samples. To assist visualization of galaxy positions at small clustercentric distances, we limit the display of projected clustercentric radius to ($r/r_{200})<3.0$.   

To aid in the separation of galaxies with small projected clustercentric radius but are infalling for the first time or are backsplash galaxies, we plot in Figure~\ref{fig:PhaseSpace} caustic profiles given by \citet{Noble2013} defined by $(\Delta v/\sigma_{v})\times (r/r_{200})=0.1$ (red solid curves) and $(\Delta v/\sigma_{v})\times (r/r_{200})=0.4$ (blue dashed curves). Using the description from \citet{Noble2013}, the region within the innermost 0.1 caustic profile (red curves) contains preferentially virialized galaxies that have fallen into the host cluster at an earlier time. For our galaxy sample, we find 15\% of the blue spirals are located in the virialized region, while 25\% of red galaxies are in this area. 

The 0.1 and 0.4 caustic profiles enclose the intermediate region that contains a mixture of galaxies that have fallen into the cluster environment early in the age of the cluster, at intermediate times, and late-time arrivals, including backsplash galaxies \citep{Noble2013}. We find 36\% of blue spirals and 43\% of red spiral galaxies are located in the intermediate region between the two caustic profiles. For galaxies located outside of the 0.4 caustic profile, we find 49\% are blue spirals and 32\% are red systems. The region outside of the 0.4 caustic profile is expected to contain mainly recently accreted galaxies \citep{Noble2013}, which we expect to contain mainly blue star-forming galaxies. 

A K-S test indicates that red and blue spirals, in terms of $(\Delta v/\sigma_{v})\times (r/r_{200})$ versus $r/r_{200}$, are statistically not selected from the same parent distribution. 

\section{Discussion}\label{sec:disc}

Our main goal for this study is to contrast and compare red and blue cluster spiral galaxies as a way to determine how blue spirals transform into red spirals without disturbing the spiral-arm structure. Galaxies have been selected in a well-defined way to minimize sample bias by including only face-on cluster spirals selected relative to the cluster red-sequence. Red-sequences where measured consistently for all clusters using the same procedure, regardless of published red-sequence fits (e.g., WINGS dataset).

Both dust and passive evolution has been suggested as a mechanism to transform normal star-forming blue spirals into optical red spirals without disturbing the spiral arm structure. \citet{Wolf2009} study based on the STAGES dataset with COMBO-17 redshifts and spectral energy distribution measurements, indicated that optically passive red spirals are undergoing star-formation, but the star formation is largely obscured due to dust in the host galaxy. Based on {\it WISE} $W2-W3$ color (see Figure~\ref{fig:WISE}), we find that 29--45\% of our red spirals are classified as ``dusty'' according to the criteria from \citet{Cui2024}. In contrast, $NUV-r$ color (Figure~\ref{fig:UV_plots}) suggests that only 1.4\% of red spirals are dusty \citep{Schawinski2007}. Even if we assume that {\it WISE} measurements provide a more robust indicator of dust in red spirals, we still need to ascertain why approximately 55--71\% of the remaining non-dusty red spirals are red at optical wavelengths.

If a significant fraction of red spirals are passively evolving, we expect that these spirals have already transitioned from blue star-forming galaxies to green valley galaxies, and will eventually transform from red passive systems into S0 galaxies \citep{Fraser-McKelvie2016,Mahajan2020,Cui2024}. Both SDSS optical colors and ultraviolet photometry support that approximately 50\% of the red spirals are passive systems (see Figures~\ref{fig:Mass_Histogram_gmr} and \ref{fig:UV_plots}). Emission line data also supports this characterization. For example, 89\% of red spirals have D\textsubscript{n}(4000)$\,>1.4$, supporting a stellar population with an age > 1 Gyr \citep{Kauffmann2003}. Also, red spirals have a lower value of both [O III] and H$\alpha$ luminosity compared to blue spirals (see Figure~\ref{fig:Luminosity_OIII_Halpha}). The plot of SFR versus galaxy mass (SFR-$M_{*}$; Figure~\ref{fig:SFR_Mass}) clearly shows that a large fraction of red spirals has a lower SFR over all measured galaxy mass compared to blue spirals. Even when normalizing the SFR with respect to galaxy mass (specific star formation rate), we find that the red spiral population has a lower sSFR than the blue galaxies (Figure~\ref{fig:sSFR_Histogram}). All of these results support the conclusion that approximately 50\% of the red spirals galaxies are passively evolving and most-likely transforming into S0 systems via secular processes.

\section{Conclusions}

We examined red and blue spiral galaxies from 115 low-redshift clusters using imaging and spectroscopic data to analyze their ultraviolet, optical, infrared, and emission-line properties. Our goal was to clarify how blue, star-forming galaxies transition into passive disk systems, with a focus on the role of red spirals in cluster environments. Infrared observations indicate that up to 45\% of optically red spirals are dusty, which can obscure star formation and result in their misclassification as passive systems. However, about half of the red spirals lack significant dust, suggesting their red color is due to passive evolution. This is supported by SDSS emission line data, including the D\textsubscript{n}(4000) spectral index, EW(H$\alpha$), EW(H$\delta$), [O III] $5007\,\text{\AA}$ luminosity, and comparisons of star formation rates with blue spirals. We conclude that red spirals are a crucial link in galaxy evolution within dense clusters and help identify the mechanisms that transform blue, star-forming galaxies into passive, red systems.
\section*{Acknowledgments}

We thank the reviewer for providing thoughtful comments and suggestions which improved the manuscript. 

This research was partially funded by ND NASA EPSCoR. We thank those students in the Department of Physics and Astrophysics at the University of North Dakota that have contributed to this research project. These students include: Haylee Archer, Jake Bartell, Darian Colgrove, Victoria Fisher, Gregory Foote, Nick Glodek, Joseph Langenwalter, Gabriel Law, Vincent Ledvina, Elijah Mathews, Walter McKee, Sydney Menne, Nicole Peterson, Evan Phillips, Carter Razink, Alexander Rice, Mason Skorup, Dean Smith, Sydney Swanson, and Benjamin Veltri.

This research has made use of the NASA/IPAC Extragalactic Database (NED) which is operated by the Jet Propulsion Laboratory, California Institute of Technology, under contract with the National Aeronautics and Space Administration. This publication makes use of data products from the Wide-field Infrared Survey Explorer \citep{Wright2010}, which is a joint project of the University of California, Los Angeles, and the Jet Propulsion Laboratory/California Institute of Technology, funded by the National Aeronautics and Space Administration. Based on observations made with the NASA Galaxy Evolution Explorer. GALEX is operated for NASA by the California Institute of Technology under NASA contract NAS5-98034.  

Funding for SDSS-III has been provided by the Alfred P. Sloan Foundation, the Participating Institutions, the National Science Foundation, and the U.S. Department of Energy Office of Science. The SDSS-III web site is http://www.sdss3.org/. SDSS-III is managed by the Astrophysical Research Consortium for the Participating Institutions of the SDSS-III Collaboration including the University of Arizona, the Brazilian Participation Group, Brookhaven National Laboratory, University of Cambridge, Carnegie Mellon University, University of Florida, the French Participation Group, the German Participation Group, Harvard University, the Instituto de Astrofisica de Canarias, the Michigan State/Notre Dame/JINA Participation Group, Johns Hopkins University, Lawrence Berkeley National Laboratory, Max Planck Institute for Astrophysics, Max Planck Institute for Extraterrestrial Physics, New Mexico State University, New York University, Ohio State University, Pennsylvania State University, University of Portsmouth, Princeton University, the Spanish Participation Group, University of Tokyo, University of Utah, Vanderbilt University, University of Virginia, University of Washington, and Yale University. 

This research made use of ds9, a tool for data visualization supported by the Chandra X-ray Science Center (CXC) and the High Energy Astrophysics Science Archive Center (HEASARC) with support from the JWST Mission office at the Space Telescope Science Institute for 3D visualization. 

This research made use of the ``K-corrections calculator'' service available at\\ http://kcor.sai.msu.ru/. 

Based on observations obtained with MegaPrime/MegaCam, a joint project of CFHT and CEA/IRFU, at the Canada-France-Hawaii Telescope (CFHT) which is operated by the National Research Council (NRC) of Canada, the Institut National des Science de l'Univers of the Centre National de la Recherche Scientifique (CNRS) of France, and the University of Hawaii.

Some/all of the data presented in this paper were obtained from the Mikulski Archive for Space Telescopes (MAST). STScI is operated by the Association of Universities for Research in Astronomy, Inc., under NASA contract NAS5-26555. Support for MAST for non-HST data is provided by the NASA Office of Space Science via grant NNX13AC07G and by other grants and contracts. 

For Tim, who to the end was interested in the advancements of space exploration.


\end{document}